# *Sand supported mixed-phase $TiO_2$ photocatalysts for water decontamination applications.*

Dorian A.H. Hanaor[(a)] and Charles C. Sorrell[(b)]

*(a)University of Sydney, School of Civil Engineering, Sydney, NSW 2006, Australia*

*(b)University of New South Wales, School of Materials Science and Engineering, Sydney, NSW 2052, Australia*

**Abstract**: Using an organometallic precursor, $TiO_2$ coatings were fabricated on surfaces of quartz, zircon and rutile sands. X-ray Diffraction, X-ray Fluorescence, UV-Vis spectroscopy and surface area measurement were used to characterise support materials. The phase composition and morphology of the coatings were characterised by laser Raman spectroscopy and SEM respectively. A packed bed reactor was used to study the inactivation of *Escherichia coli* in recirculating water by the supported photocatalysts. It was found that the sand grains were well coated with a homogenous layer of $TiO_2$ and coatings were well adhered, exhibiting a mixed anatase-rutile composition after firing at 850°C. Photocatalytic activity was highest in coatings applied to quartz sand, although sterilisation of the recirculating water was not achieved with any of the materials investigated. The advantages of quartz as a $TiO_2$ photocatalyst support material are likely a result of this material's higher purity and optical transmittance. Potential enhancement through Si doping cannot be ruled out.

## 1. Introduction

$TiO_2$ photocatalysts are of increasing interest owing, in part, to their potential applications in solar-driven environmental remediation. In recent decades $TiO_2$ has been shown to facilitate the decomposition of organic pollutants and pesticides and the inactivation of bacteria in water through the formation of adsorbed radicals by photogenerated electron-hole pairs at the photocatalyst surface [1-3].

The anatase phase of $TiO_2$ is the first phase formed through most synthesis routes and undergoes rutilation through thermal treatment [4]. Rutile, which exhibits a lower band gap relative to anatase, generally exhibits lower effective surface area and consequently, lower photocatalytic activity than anatase [3-5]. It has been widely reported that anatase-rutile mixtures exhibit enhanced photocatalytic activity, relative to single phase $TiO_2$, through enhanced electron-hole separation (improved exciton lifetime), allowing higher levels of radicals to be generated [6-9].

Water purification by semiconductor photocatalysis has most commonly been studied using suspensions of $TiO_2$ powders in treated water [10-12]; this approach entails a catalyst recovery step to remove dispersed powders from treated water, rendering such systems unsuitable for significant upscaling. For large capacity water purification, the use of immobilised $TiO_2$ on catalyst supports of high surface area is an attractive approach for eliminating the requirement of complex liquid/powder separation processes.

Various catalyst support materials of increased surface area have been utilised to enhance the available surface area of immobilised $TiO_2$ photocatalysts for water purification applications.

These support materials include glass fibres [13, 14], porous silica [15], silica gel [16], amorphous silica nano-particles [17], carbon particles [18, 19], and zeolites [20].

Sands are attractive as enhanced-surface-area catalyst support materials for large scale water purification as a result of their low costs and the ease with which they can be separated from water. Through differing compositions and structures, different types of sand as catalyst supports may enhance or impair photocatalytic activity of $TiO_2$ immobilised on the sand grain surfaces.

The present work investigates the fabrication of nano-crystalline mixed-phase $TiO_2$ coatings on quartz, zircon and rutile sand for use in water purification. $TiO_2$ immobilised on quartz sand ($SiO_2$) was investigated as such materials may exhibit enhanced photoactivity caused by the diffusion of Si atoms from the underlying sand into the $TiO_2$ coating. Doping by Si has been reported to improve the performance of $TiO_2$ photocatalysts in water purification applications [21-23]. This has been reported to be a result of improved charge carrier separation at $TiO_2$-$SiO_2$ interfaces.

Immobilised $TiO_2$ on zircon sand ($ZrSiO_4$) was studied in the present work as this system has the potential to impart enhanced photocatalytic performance through the nucleation of elongated prismatic rutile crystallites on zircon surfaces alongside residual anatase as reported previously [24]. Further, diffusion of Zr from the underlying sand may enhance performance as suggested by reports that Zr doping enhances surface area and photoactivity of $TiO_2$ [25, 26].

Finally, the use of rutile sand for the immobilisation of $TiO_2$ may improve charge carrier separation through the previously reported synergism between large rutile grains and fine anatase particles in the coating [9, 27]

## 2. *Experimental*

Glassmaking quartz sand and naturally occurring rutile and zircon sands (Wallarah Minerals, Doyalson North, Australia) were used as support materials. 50 gram samples of sand were ultrasonically cleaned successively in acetone and ethanol, and subsequently dried in air by heating at 110°C. The crystallography of the sands was examined by X-ray diffraction (XRD) carried out using a Phillips X'pert MPD XRD system with Cu Kα X-ray emission. Surface area of the sands was evaluated by BET analysis of $N_2$ adsorption isotherms using a Micromeritics ASAP 2020 Physisorption Analyzer. Impurity levels in the sands were analysed by wavelength dispersive X-ray fluorescence (XRF) using a Philips PW2400 XRF spectrometer.

A coating method using an organometallic derived colloidal suspension was employed. In this method, subsequent to cleaning, the sands were added gradually to a solution of 0.5M titanium tetra-isopropoxide, TTIP (97%, Sigma Aldrich) in 50 ml of isopropanol (100%, Univar), a commonly used solvent for TTIP. This was carried out under rapid stirring by a magnetic stirrer. By the dropwise addition of water diluted to 10 mL with the appropriate volume of isopropanol, the rapidly stirring slurry of sand in TTIP solution was hydrolysed at a hydrolysis ratio of R=4, giving a ratio of $H_2O$:TTIP of 4:1. The use of this moderately low hydrolysis ratio was chosen in order to avoid the formation of large and dense agglomerates that would not be able to adhere to sand surfaces. The additional isopropanol and dropwise addition were used for the purpose of moderating the rapidity of the hydrolysis and condensation reactions. After ~5 minutes of continued stirring colloidal particles of amorphous $TiO_2$ were formed in similarity to earlier work [28], and deposited onto sand surfaces.

Coatings on the different types of sands were crystallised and adhered by firing in air using a muffle furnace at 850 °C with heating /cooling rates of 2° $min^{-1}$. These firing parameters were chosen as they were found to yield a mixed phase composition of $TiO_2$ coatings on single crystal quartz substrates in previous work [29]. The slow rates of heating / cooling were chosen in order to reduce spalling.

Laser Raman microspectroscopy was carried out in conjunction with optical microscopy to ascertain the phase composition of the coatings using an InVia Raman Microspectrometer with laser excitation at 514 nm. Scanning electron microscopy (SEM) was carried out using Hitachi s900 and s3400 microscopes in order to examine the quality of $TiO_2$ coatings on the sand particles and the morphologies of $TiO_2$ grains within these coatings.

The photocatalytic performance of the supported $TiO_2$ materials prepared in the presesnt work was evaluated by studying the inactivation of *Escherichia coli* (*E. coli*) AN180 (School of Biotechnology and Biomolecular Sciences, UNSW, Australia), using the bactericidal reactor illustrated in **Fig. 1.** The apparatus was sterilised by recirculating 80 vol% ethanol and subsequently the system was flushed with autoclave-sterilised deionized water. A reservoir of 1L of sterilised-deionized water was used for recirculation. Recirculating water was inoculated with 1 ml of overnight-incubated *E. coli* culture in Tryptone Soy Broth (Oxoid, Basingstoke, UK). Recirculating water flowed through a packed bed of 40g of coated sand which was irradiated over a length of 350mm and width of 20mm placed 50mm below a 15 W UV light source with an emission peak at 350 nm (Sylvania BL350) giving an irradiance level of ~3.7 $Wm^{-2}$ at the catalyst surface as determined using a Digitech QM 1587 light-meter.

Inactivation of the bacteria in the recirculating water was examined by determining the concentration of viable colony forming units (CFUs) in the recirculating water according to ASTM D5465 [30]. 1 ml aliquots of water were taken at fixed time intervals and serially diluted at 1:9 ratios in sterile 0.1% peptone water (Oxoid). Subsequently 100 µl aliquots of the appropriate dilutions were spread-plated on Tryptone Soy Agar gel plates (Oxoid) and these were incubated for 24 hours at 37 °C. Subsequent to incubation, colonies were enumerated and the concentration of colony forming units (CFUs) in the reactor water was determined.

To evaluate the penetration of UV irradiation through the different types of sand in the packed bed reactor, UV-Vis transmission, through 1 mm layers of sand, was measured using a Perkin Elmer Lambda-35 UV-Visible spectrometer, with grains of sands allowed to settle under gravity in the measurement cuvettes.

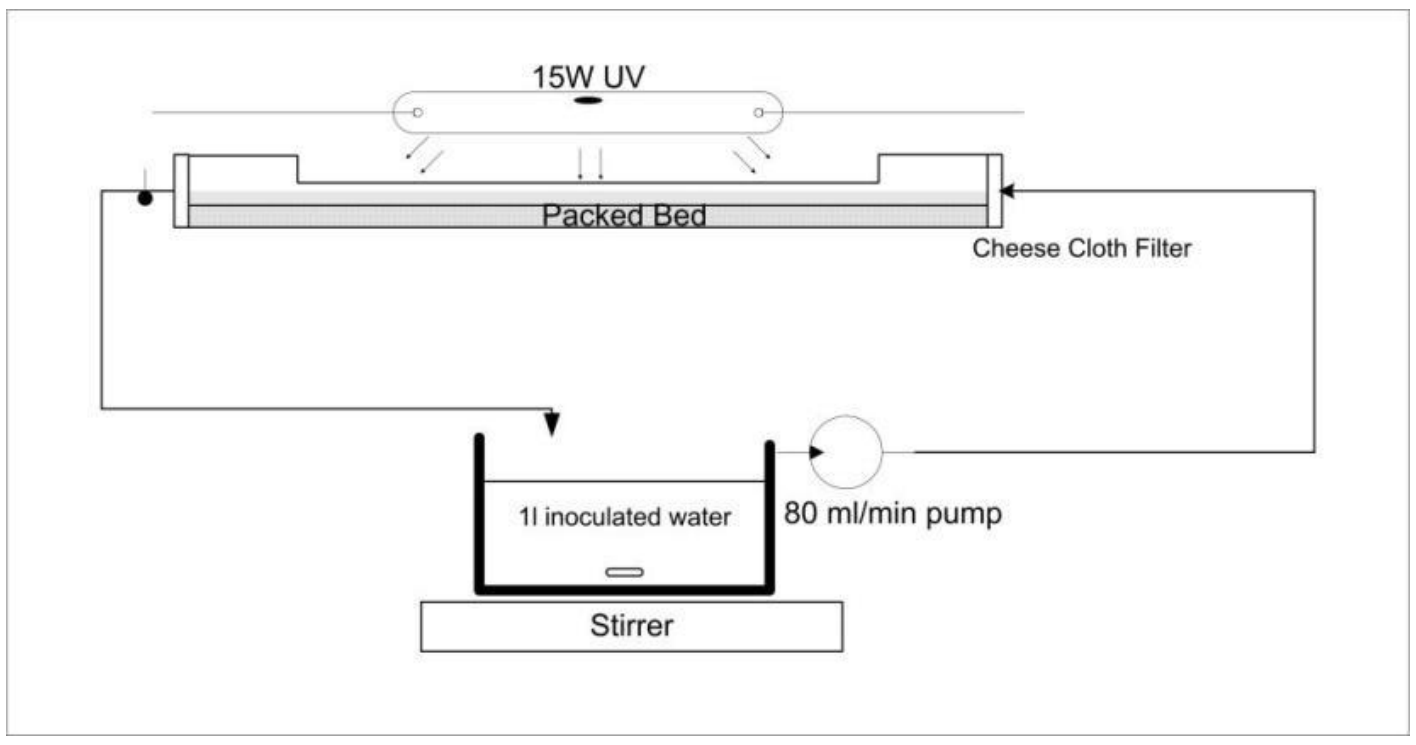

**Fig.1. Diagram of bactericidal reactor**

## 3. *Results*

### 3.1. Characterisation of support materials

XRD patterns of sands are shown in **Fig.2.** Quartz sand is of the α-quartz structure, space group $P3_121$ Zircon sand exhibits a tetragonal structure, space group $I4_1/amd$ and Rutile sand exhibits a tetragonal structure, space group P4/mnm [31]. No secondary crystalline phases were identified in the sands by XRD.

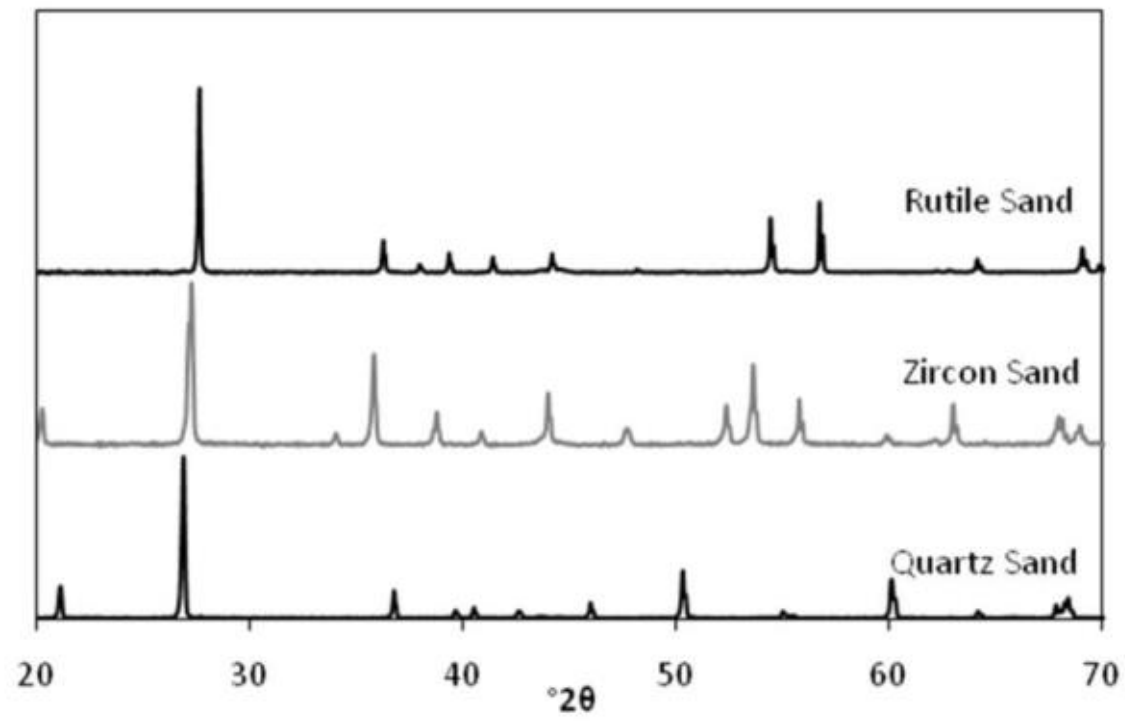

**Fig.2. XRD patterns of sands used as catalyst support materials**

The properties of the sands used as catalyst support materials are summarised in **Table 1.** It can be seen that quartz sand exhibits a lower surface area in comparison with zircon and rutile sands. This is a natural consequence of the larger sizes, smoother surfaces and more spherical morphologies of grains of this type of sand. While no secondary crystalline structures were apparent through XRD analysis, X-Ray Fluorescence (XRF) shows significant levels of impurities in zircon and rutile sands. The levels of impurities as determined by XRF show rutile sand has the highest levels of impurities while quartz sand exhibits low levels of impurities. This was expected as the quartz sand used in this work is of glassmaking specification.

**Table 1. Summary of support materials used in the present work.**

| **Sand** | **Formula** | **Average Grain size (μm)** | **Surface area (BET) $m^2g^{-1}$** | **Crystal structure** | **Impurites (wt%)** |
| --- | --- | --- | --- | --- | --- |
| Quartz | $SiO_2$ | ~250 | 0.019 | Trigonal | $Al_2O_3$: 0.03%<br>$ZrO_2$ 0.05%<br>$TiO_2$: 0.04%<br>$Fe_2O_3$: 0.01% |
| Zircon | $ZrSiO_4$ | ~120 | 0.248 | Tetragonal | $HfO_2$: 1.54%<br>$Al_2O_3$: 0.56%<br>$TiO_2$: 0.24%<br>$Fe_2O_3$: 0.13% |
| Rutile | $TiO_2$ | ~100 | 0.105 | Tetragonal | $SiO_2$: 1.76%<br>$ZrO_2$: 0.73%<br>$Fe_2O_3$: 0.42%<br>$Al_2O_3$: 0.24%<br>$Na_2O$: 0.20% |

UV-Vis transmission spectroscopy of the sands used as photocatalyst support materials, was carried out to ascertain the penetration of UVA and visible light through the packed bed of sand used in the bactericidal reactor. The UV-Vis transmission through 1 mm of quartz sand was significantly higher than the transmission through a similar profile of the darker zircon and rutile sands.

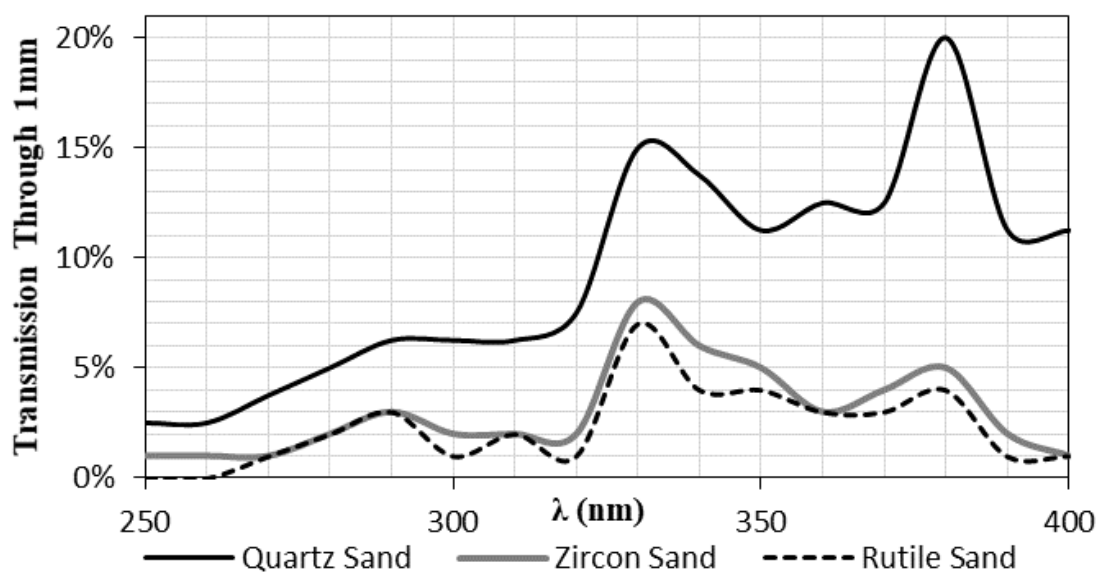

**Fig.3. UV-Vis transmission through 1 mm of the sands used in this work**

### 3.2. Coating morphology

Coatings on all types of sand were consistent in appearance, and coating thicknesses varied in the region of ~30-300 nm, with an average coating thickness of ~200 nm. The typical appearance of coatings is shown in **Fig. 4.** Localised regions exhibiting the absence of $TiO_2$ coating as a result of spalling were observed in rutile sand, as shown in **Fig. 5.,** however this was an infrequent observation and overall coatings were well-adhered. The microstructure of the coatings consisted of $TiO_2$ grains of 30-50nm as shown in **Fig. 6.** Cracking was prevalent to a similar extent in all coatings. As shown in **Fig. 7**, SEM images of unfired coated sand showed that cracking was present subsequent to drying rather than formed in the process of firing. The coatings were resistant to prolonged use in the recirculating bactericidal reactor and no significant deterioration in coatings was evident subsequent to use in the reactor shown in Fig. 1

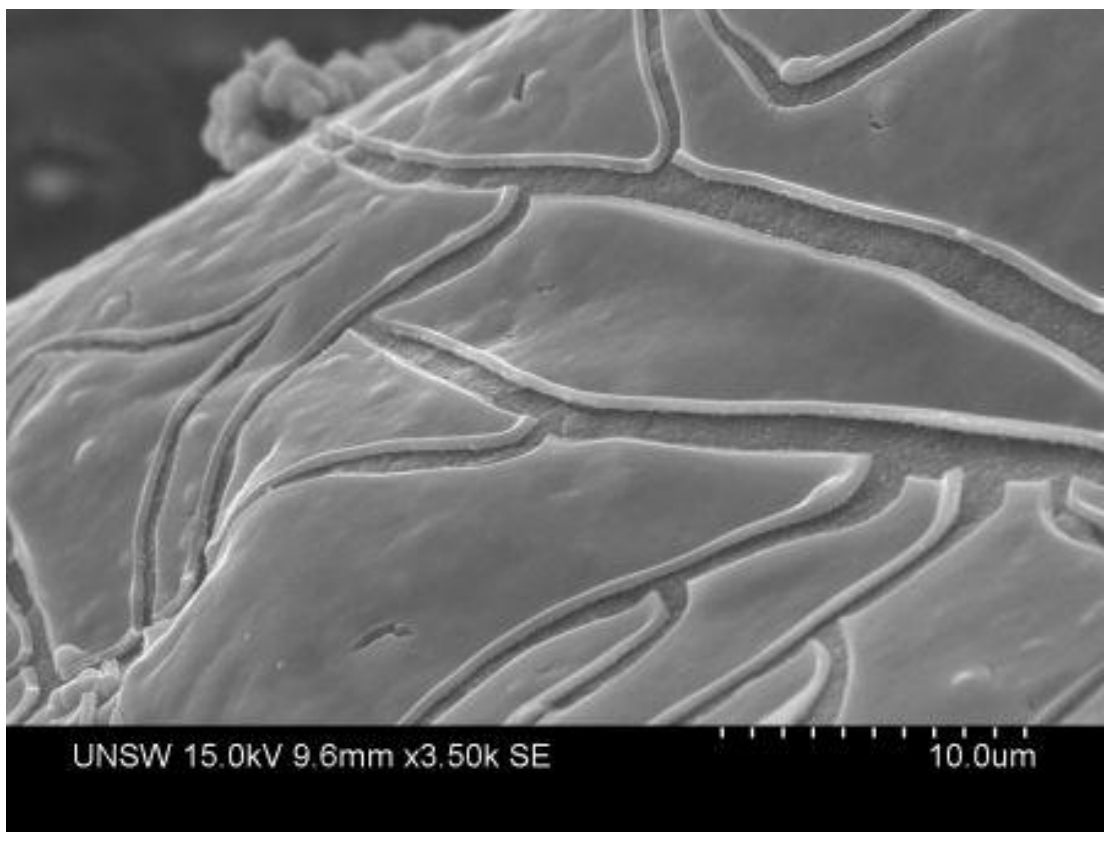

**Fig. 4. Sub-micron thick coating of mixed anatase-rutile $TiO_2$ on rutile sand. Morphology shown is typical for all types of sand**

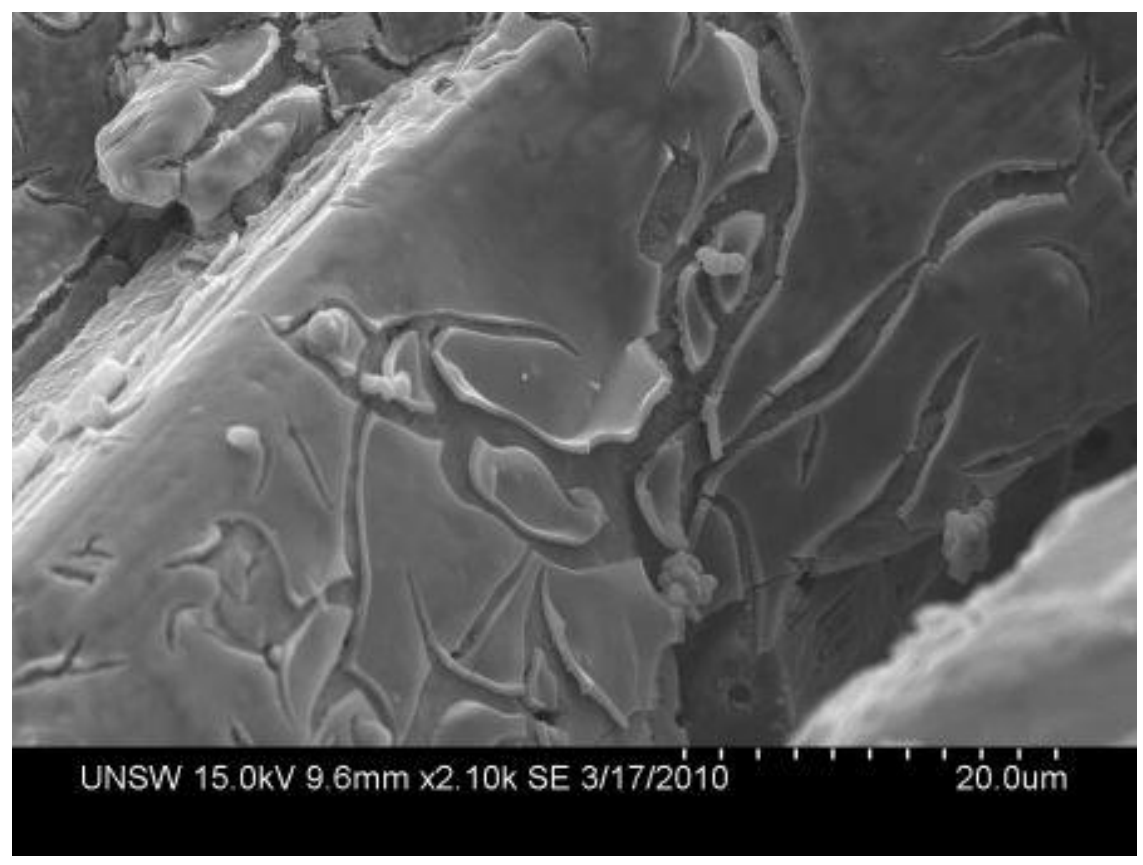

**Fig. 5. Region of spalling $TiO_2$ coating on sand (rutile) subsequent to firing at 850°C**

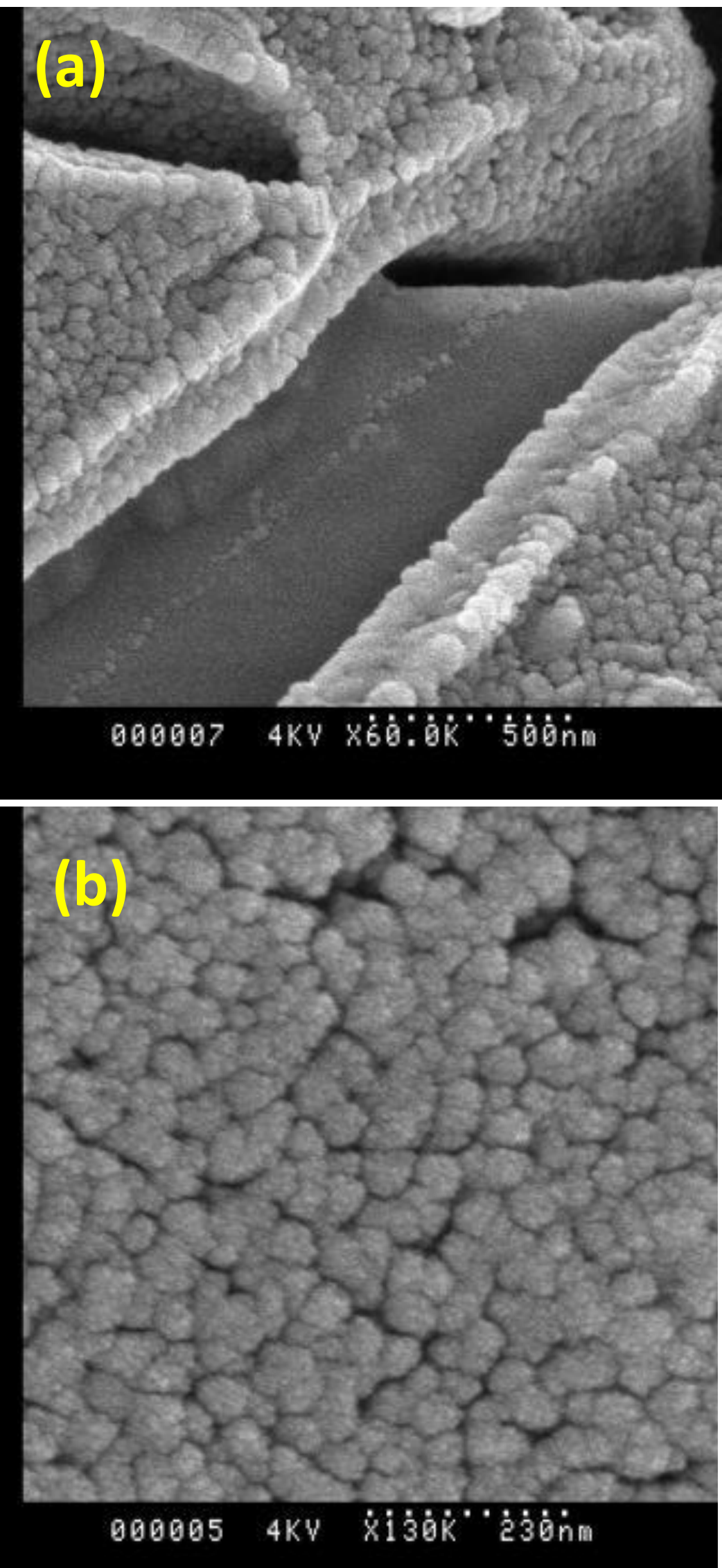

**Fig.6. Microstructure of $TiO_2$ coatings on sand (quartz) (a) Cracked coating interface (b) Coating micrograph**

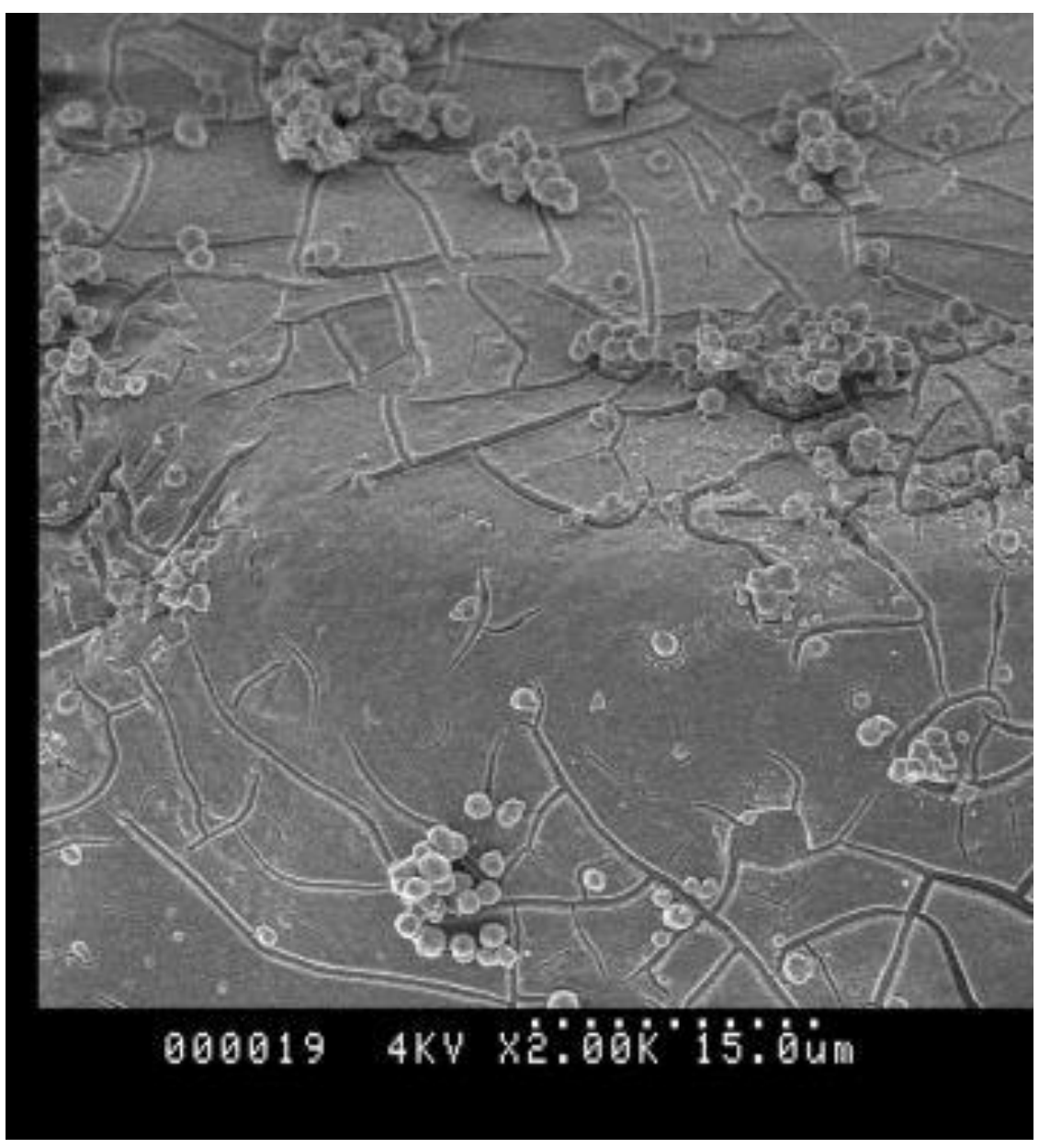

**Fig. 7. Unfired $TiO_2$ coated sand (Zircon), with excess loosely adhered $TiO_2$ particles visible on the surface**

### 3.3. Phase composition of coatings

XRD analysis was deemed to be unsuitable for phase analysis of coatings as the peaks from the grains of sand are significantly more dominant in such analyses relative to peaks from the thin nanocrystalline coatings. For this reason laser Raman microspectroscopy was used to differentiate the phases present in coatings using similar parameters to those reported elsewhere [32]. Alongside varying significant levels of peaks from the underlying and partly exposed sands, Raman spectra of coated sands fired in air at 850°C, shown in **Fig. 8**, show biphasic anatase + rutile compositions in the coatings with anatase peaks being dominant (in part, a result of the stronger Raman shift by this phase). The rutile Raman shifts present in coatings on zircon and quartz sands are evident in the shoulders of the anatase peaks 395 and at 640 $cm^{-1}$ [32]. The rutile Raman shift in coatings on rutile sands is strong as a result of the combined effects from a rutile phase in the coating and from the underlying support material. The Raman spectra shown here are not appropriate for quantitative analysis, however it is likely that anatase is present in a larger quantity than rutile, owing to the weak Raman shift exhibited by rutile in these spectra.

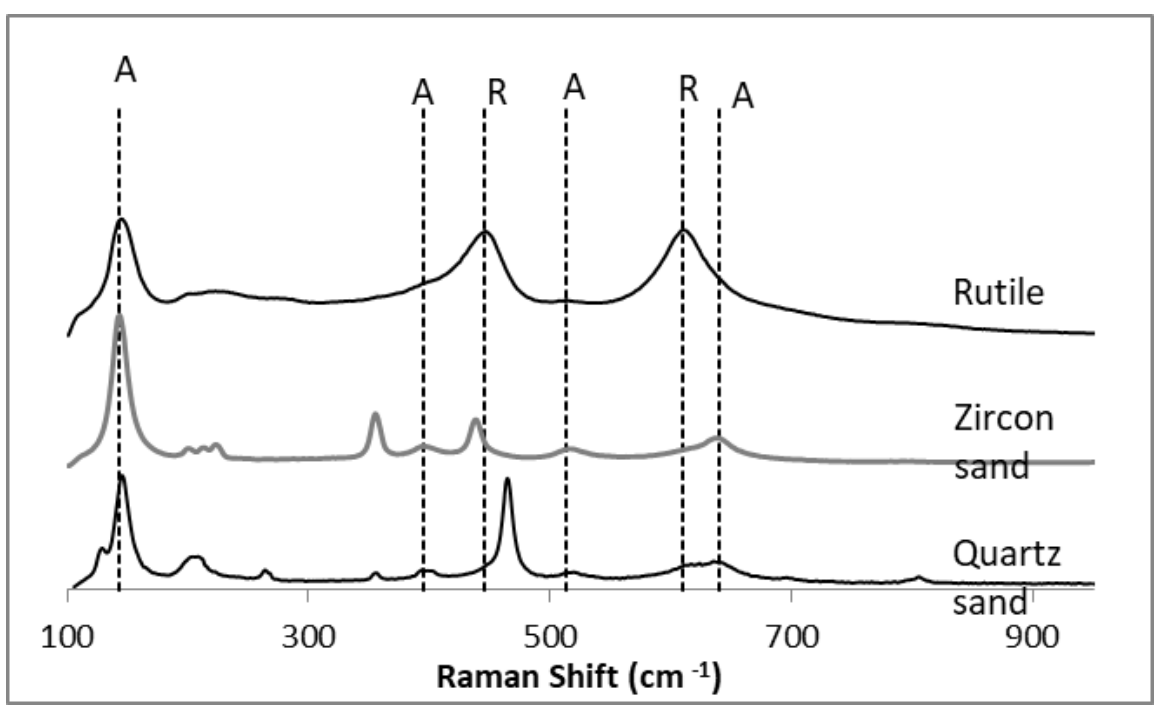

**Fig. 8. Raman spectra of $TiO_2$ coatings on grains of different sand types**

### 3.4. Bactericidal activity

The photocatalyzed inactivation of bacteria is shown in **Fig. 9.** In order to determine the background bactericidal activity in the absence of a photocatalyst, uncoated quartz sand was used in the reactor with and without UV irradiation, in the UVA region (325-400 nm, emission peak at 350 nm). It can be seen that the inactivation of *E. coli* in the absence of $TiO_2$ coatings is negligible even in the presence of UV irradiation. This is a result of the longer wavelength of the UV source used in comparison with UV irradiation typically used in antimicrobial lamps [33]. In the presence of $TiO_2$ coated sand, inactivation of bacteria under UV irradiation is evident, being significantly more pronounced in the coated quartz sand. No complete sterilisation of the recirculating water was achieved within the timeframe of the experiments, rather the concentration of *E. coli*, as determined by counting colony forming units (CFUs), diminished to a level at which the bactericidal activity of the photocatalyst was in effect offset by the natural multiplication of the bacteria in the water.

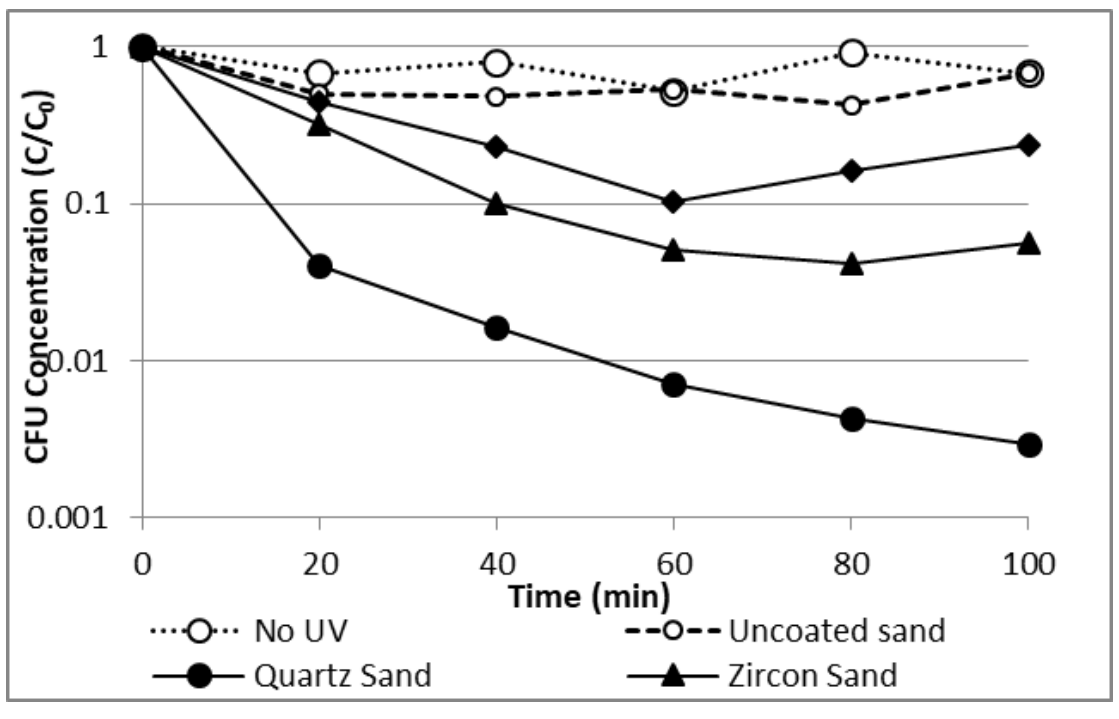

**Fig. 9. Bactericidal activity of coated sands as determined by CFU counts**

## 4. Discussion

### 4.1. Coating morphology

Using a colloidal-sol deposition method, sand grains exhibited reasonably homogenous coverage by $TiO_2$ coatings as shown in Fig. 4. A reaction between surface adsorbed hydroxyl groups at the sand surfaces and the titanium tetra-isopropoxide precursor may result in the formation of (Si/Ti/Zr)-O-Ti bonding and further serve to enhance coating integrity and adhesion. Cracking was prevalent in all coatings and SEM micrographs of unfired material, such as that shown in Fig. 7, have shown that this crack formation occurs during drying, likely as the result of shrinkage. Although most grains were well coated, infrequently occurring areas devoid of $TiO_2$ were observed, particularly in coatings on rutile sands. Spalling of $TiO_2$ coatings may have occurred as a result of differential shrinkage in combination with localised poor adhesion due to remaining contaminants. As rutile sand was of the lowest purity, poor coating adhesion on this material may have been the result of the combustion of residual organic material and/or other impurities on the sand surfaces. The nanocrystalline microstructure of $TiO_2$ grains resembles photocatalytic films synthesised on quartz substrates and subjected to a similar thermal treatment [29]. Overall coating adhesion was good on all three support materials with little deterioration of the coating evident after use in the bactericidal reactor. It is likely that localised poor adhesion allows a limited degree of coating removal through spalling and abrasion, however the results indicate that with appropriate synthesis parameters, adequately durable coatings are achievable to allow prolonged lifetime in a similar reactor, providing similarly low levels of support-material-agitation are present.

#### 4.1.1. Phase Composition

As indicated by Raman spectra, $TiO_2$ coatings exhibited a mixed phase composition after firing at 850 °C. The phase transformation behaviour of these coatings is similar to what has been observed in coatings fabricated on single crystal quartz substrates [29]. The anatase to rutile phase transformation is inhibited in coatings on the sands used in this work in comparison with unsupported anatase which is generally reported to transform to rutile at temperatures around 600°C [4]. The stabilisation of the anatase phase in coatings on sand is likely a result of the diffusion of Si, Zr and Al into the $TiO_2$ coatings. These elements are reported to inhibit the formation of rutile through the restriction of the atomic rearrangement involved in the phase transformation. [23, 34-37] The use of lower firing temperatures for similar coatings may result in similar mixed phase coatings if anatase-to rutile transformation promoting dopants, such as V, Co and Ni , are added [38]. However the use of lower firing temperatures may impede the formation of good coating adhesion. Higher temperatures are likely to improve adhesion, however these are likely to result in single phase rutile coatings exhibiting poor performance resulting from crystal growth. The use of a so-gel micelle nanostructure precursor as reported elsewhere[39] may be applicable in order to facilitate the use of high temperature adhesion processes while maintaining nanostructured coatings.

#### 4.1.2. Photocatalytic Performance

From studying the inactivation of *E. coli* in recirculating water it is apparent that quartz sand is a superior support material despite exhibiting a lower surface area as determined by BET analysis. This may be the result of higher purity and thus lower levels of contamination of the photocatalyst by secondary phases which could act as charge carrier recombination centres. Rutile sand, and to a lesser extent zircon sand, exhibited significant levels of iron impurities as shown by XRF analysis. Under certain conditions, Fe, and other transition metals [40-43] and alkali impurities, [44-46] have been reported to impair photocatalytic activity and it is possible the presence of such impurities in rutile and zircon sands contributed to the poor bactericidal performance of $TiO_2$ coatings supported on these materials. It should be noted that the presence of very low levels of transition metal dopants/impurities, such as the levels observed in quartz sand by XRF, has been reported as being beneficial for photoactivity. [47]

Diffusion of Si from the quartz sand into the $TiO_2$ lattice may enhance photoactivity in similarity to observations reported elsewhere [21-23]. While this effect is likely also in zircon sand and in rutile sand (owing to the $SiO_2$ impurities in this material), it is possible that the levels of diffusion from quartz are greater, and/or the beneficial effects of Si doping are outweighed by detrimental levels of transition and alkali metal impurities. Conclusive evidence of the presence of Si in the coatings was not achieved as XRD, Raman and other spectroscopic techniques are not sensitive enough to ascertain with confidence whether interstitial Si is present in the $TiO_2$ lattice.

Further impairing the performance of rutile and zircon sands in the bactericidal reactor is the poor UV-Vis transmittance of these materials, as shown in Fig. 3. This leads to low penetration of the UV irradiation into the packed bed and thus lower levels of irradiated $TiO_2$ coated surfaces.

A packed bed reactor such as the one used in this work suffers the drawback of poor catalyst exposure as only the top layer of grains is exposed directly to the UV source. This demonstrative reactor was chosen due to its simplicity and the ease with which experiments could be repeated. A more effective reactor would involve greater levels of coated surfaces subjected to impingent UV irradiation. The low cost of the raw materials used in the present work mean that large scale fabrication can be undertaken inexpensively.

## 5. *Conclusions*

Thin well-adhered coatings of nanocrystalline $TiO_2$ can be fabricated on various types of sand enabling the cheap production of an enhanced-surface area supported photocatalyst.

The anatase to rutile phase transformation is inhibited in $TiO_2$ supported on sand grains as a result of the diffusion of structure-stuffing cations, which restrict the atomic rearrangement involved in the formation of rutile from anatase.

The use of glass-making quartz sand as a photocatalyst support material appears to bring about superior photoactivity in comparison with naturally occurring zircon and rutile sands, likely as a result of higher purity and optical transmittance.

**Acknowledgments**

The authors acknowledge access to the UNSW node of the Australian Microscopy and Microanalysis Research Facility (AMMRF) and the spectroscopy lab at the Mark Wainwright analytical centre at UNSW. We further acknowledge the support and assistance of the Institute of Materials and Engineering Sciences at the Australian Nuclear

Science and Technology Organisation (ANSTO) and the Australian Institute of Nuclear Science and Engineering (AINSE).

## *References*

[1] C. Dominguez, J. Garcia, M. Pedraz, A. Torres, M. Galan, *Catalysis today* **1998**, *40*, 85.
[2] R. W. Matthews, *Journal of Physical Chemistry* **1987**, *91*, 3328.
[3] A. Mills, R. H. Davies, D. Worsley, *Chemical Society Reviews* **1993**, *22*, 417.
[4] D. Hanaor, C. C. Sorrell, *Journal of Materials Science* **2011**, *46*, 855.
[5] A. Sclafani, J. M. Herrmann, *Journal of Physical Chemistry* **1996**, *100*, 13655
[6] D. C. Hurum, A. G. Agrios, K. A. Gray, T. Rajh, M. C. Thurnauer, *J. Phys. Chem. B* **2003**, *107*, 4545.
[7] S. Mahshid, M. Askari, M. Sasani Ghamsari, N. Afshar, S. Lahuti, *Journal of Alloys and Compounds* **2009**, *478*, 586.
[8] D. C. Hurum, K. A. Gray, T. Rajh, M. Thurnauer, *Journal of Physical Chemistry B* **2005**, *109*, 977.
[9] T. Ohno, K. Tokieda, S. Higashida, M. Matsumura, *Applied Catalysis A* **2003**, *244*, 383.
[10] P. C. Maness, S. Smolinski, D. M. Blake, Z. Huang, E. J. Wolfrum, W. A. Jacoby, *Applied and Environmental Microbiology* **1999**, *65*, 4094.
[11] V. Luca, M. Osborne, D. Sizgek, C. Griffith, P. Z. Araujo, *Chem. Mater* **2006**, *18*, 6132.
[12] G. P. Fotou, S. Vemury, S. E. Pratsinis, *Chemical engineering science* **1994**, *49*, 4939.
[13] J. C. Ireland, P. Klostermann, E. W. Rice, R. M. Clark, *Applied and Environmental Microbiology* **1993**, *59*, 1668.
[14] K. Hofstadler, R. Bauer, S. Novalic, G. Heisler, *Environmental Science & Technology* **1994**, *28*, 670.
[15] R. Van Grieken, J. Aguado, M. J. Lopez-Muʊoz, J. Marugon, *Journal of Photochemistry & Photobiology, A: Chemistry* **2002**, *148*, 315.
[16] Z. Ding, X. Hu, G. Q. Lu, P. L. Yue, P. F. Greenfield, *Langmuir* **2000**, *16*, 6216.
[17] S. Karino, J. Hojo, *Journal of the Ceramic Society of Japan* **2010**, *118*, 591.
[18] E. Carpio, P. Zuniga, S. Ponce, J. Solis, J. Rodriguez, W. Estrada, *Journal of Molecular Catalysis. A, Chemical* **2005**, *228*, 293.
[19] Y. Li, X. Li, J. Li, J. Yin, *Catalysis Communications* **2005**, *6*, 650.
[20] Y. H. Hsien, C. F. Chang, Y. H. Chen, S. Cheng, *Applied Catalysis B, Environmental* **2001**, *31*, 241.
[21] H. R. Jafry, M. V. Liga, Q. Li, A. R. Barron, *Environmental Science & Technology*.
[22] C. Anderson, A. J. Bard, *The Journal of Physical Chemistry* **1995**, *99*, 9882.
[23] D. M. Tobaldi, A. Tucci, A. S. Skapin, L. Esposito, *Journal of the European Ceramic Society* **2010**, *30*, 2481.
[24] D. A. Hanaor, W. Xu, M. Ferry, C. C. Sorrell, *Journal of Crystal Growth* **2012**, *359*, 83.
[25] J. Lukac, M. Klementová, P. Bezdicka, S. Bakardjieva, J. Subrt, L. Szatmáry, Z. Bastl, J. Jirkovsk *Applied Catalysis B: Environmental* **2007**, *74*, 83.
[26] Y. M. Wang, S. W. Liu, M. K. Lu, S. F. Wang, F. Gu, X. Z. Gai, X. P. Cui, J. Pan, *Journal of Molecular Catalysis A: Chemical* **2004**, *215*, 137.
[27] V. Loddo, G. Marc, C. Mart n, L. Palmisano, V. Rives, A. Sclafani, *Applied Catalysis B, Environmental* **1999**, *20*, 29.
[28] D. A. Hanaor, I. Chironi, I. Karatchevtseva, G. Triani, C. Sorrell, *Advances in Applied Ceramics* **2012**, *111*, 149.
[29] D. Hanaor, G. Triani, C. C. Sorrell, *Surface and Coatings Technology* **2011**, *205*, 3658.
[30] ASTM, in *ASTM D5465 Standard Practice for Determining Microbial Colony Counts from Waters Analyzed by Plating Methods*, Philadelphia, Pa **2004**.
[31] FIZ Karlsruhe, **1997**.
[32] J. L. Gole, S. M. Prokes, O. J. Glembocki, *The Journal of Physical Chemistry C* **2008**, *112*, 1782.
[33] S. H. Shah, *Water, Air, & Soil Pollution: Focus* **2009**, *9*, 529.
[34] M. Hirano, N. Nakahara, K. Ota, O. Tanaike, N. Inagaki, *Journal of Solid State Chemistry* **2003**, 39
[35] M. K. Akhtar, S. E. Pratsinis, S. V. R. Mastrangelo, *Journal of the American Ceramic Society* **1992**, *75*, 3408.
[36] D. J. Reidy, J. D. Holmes, M. A. Morris, *Ceramics International* **2006**, *32*, 235.
[37] K. Okada, N. Yamamoto, Y. Kameshima, A. Yasumori, *Journal of the American Ceramic Society* **2001**, *84*, 1591.
[38] J. Choi, H. Park, M. R. Hoffmann, *The Journal of Physical Chemistry C* **2009**, *114*, 783.
[39] A. Ogden, J. A. Corno, J.-I. Hong, A. Fedorov, J. L. Gole, *Journal of Physics and Chemistry of Solids* **2008**, *69*, 2898.
[40] J. C. Colmenares, M. A. Aramendia, A. Marinas, F. J. Urbano, *Applied Catalysis A* **2006**, *306*, 120.
[41] K. Wilke, H. Breuer, *Journal of Photochemistry and Photobiology A: Chemistry* **1999**, *121*, 49.
[42] D. Dvoranova, V. Brezova, M. Mazur, M. Malati, *Applied Catalysis B: Environmental* **2002**, *37*, 91.
[43] A. Di Paola, G. Marci, L. Palmisano, M. Schiavello, K. Uosaki, S. Ikeda, B. Ohtani, *The Journal of Physical Chemistry B* **2002**, *106*, 637.
[44] A. Fernandez, G. Lassaletta, V. M. Jimenez, A. Justo, A. R. Gonzalez-Elipe, J. M. Herrmann, H. Tahiri, Y. Ait-Ichou, *Applied Catalysis B, Environmental* **1995**, *7*, 49.
[45] J. M. Herrmann, *Catalysis Today* **1999**, *53*, 115.
[46] J. C. Yu, J. Yu, J. Zhao, *Applied Catalysis B: Environmental* **2002**, *36*, 31.
[47] A. Linsebigler, G. Lu, J. T. Yates, *Chem. Rev.* **1995**, *95*, 735